\documentclass{ws-ijmpa}
\usepackage{amssymb,graphicx,amsmath,graphics,subfigure,xfrac,url,xcolor}

\newcommand\D{\mathrm{d}}

\begin{document}

\title{Study of the Superradiance Phenomenon in the $\alpha$--attractor Potential using the Log Derivative Method}

\author{\'Angel Salazar}
\address{Yachay Tech University, School of Physical Sciences and Nanotechnology, Hda. San Jos\'e S/N y Proyecto Yachay, 100119, Urcuqu\'i, Ecuador.}

\author{Quray Potos\'i}
\address{Yachay Tech University, School of Physical Sciences and Nanotechnology, Hda. San Jos\'e S/N y Proyecto Yachay, 100119, Urcuqu\'i, Ecuador.}

\author{David Laroze}
\address{Instituto de Alta Investigación, Universidad de Tarapacá, Casilla 7 D,
1000000--Arica, Chile.
}

\author{Laura M. P\'erez}
\address{Departamento de Ingeniería Industrial y Sistemas, Universidad de
Tarapacá, Casilla 7 D, 1000000--Arica, Chile.
}

\author{Benjamin De Zayas}
\address{Yachay Tech University, School of Physical Sciences and Nanotechnology, Hda. San Jos\'e S/N y Proyecto Yachay, 100119, Urcuqu\'i, Ecuador.}

\author{Clara Rojas}
\address{Yachay Tech University, School of Physical Sciences and Nanotechnology, Hda. San Jos\'e S/N y Proyecto Yachay, 100119, Urcuqu\'i, Ecuador.\\
crojas@yachaytech.edu.ec}

\maketitle

\begin{abstract}
In this article, we solved the time--independent one--dimensional Klein--Gordon equation in the presence of $\alpha$--attractor potential  using the Log derivative method. We calculated the reflection  coefficient $\mathcal{R}$ and the transmission coefficient $\mathcal{T}$, showing that the superradiance phenomenon is present. In order to demonstrate the accuracy of our method, we performed a comparison with the analytical solution for the hyperbolic tangent potential.

\end{abstract}


\section{Introduction}	

The study of relativistic scattering remains a useful framework for understanding how quantum systems behave in the presence of external fields. In particular, the Klein--Gordon equation provides one of the simplest relativistic descriptions of spin--0 particles \cite{klein:1926,gordon:1926}, and it has been widely used to explore different types of interactions and potential models \cite{valladares2023superradiance,badalov:2023,benchiheub2024,ikhdair:2010}. Even in one spatial dimension, the introduction of a nontrivial potential can lead to behaviors that are not immediately evident. 

A common approach is to consider smooth potential barriers since they allow one to interpolate between asymptotic regions without introducing discontinuities. The hyperbolic tangent potential is a well--known example of this type of model. It has been studied extensively in the literature \cite{badalov:2023,rojas:2014,fernandez:2016,ahmadov:2021}, mainly because it captures the essential features of a barrier while still being analytically manageable. In fact, its solutions can be written in terms of special functions \cite{abramowitz:1965}, which makes it particularly useful as a reference case.

In relativistic settings, one of the features that emerge naturally is superradiance \cite{manogue1988klein}. This effect, which can be understood as a reflection coefficient larger than unity, is closely related to the Klein paradox \cite{manogue1988klein,dombey1999seventy}. Although it may seem counterintuitive at first, it is a direct consequence of the structure of the relativistic dispersion relation. More recently, similar behavior has been reported in different models, suggesting that superradiance is not restricted to a specific potential \cite{valladares2023superradiance}.

Recent analytical investigations into the $\alpha$--attractor potential \cite{salo:2021}, have demonstrated a fundamental obstruction to its integrability. Using Picard--Vessiot theory, it has been established that the differential Galois group associated with this system is isomorphic to $SL(2,\mathbb{C})$. Since this group is non--solvable, it implies the non--existence of Liouvillian solutions\cite{liuvilliana:2026}. Furthermore, the intrinsic transcendence of the potential prevents its reduction to a form defined over a rational function field $\mathbb{C}(z)$, effectively excluding the possibility of expressing solutions as compositions of special functions such as hypergeometric, Whittaker, Bessel, or Heun functions\cite{liuvilliana:2026}. Consequently, the study of its scattering properties necessitates the use of robust numerical frameworks.

From a computational perspective, obtaining reflection and transmission coefficients is not always straightforward. Analytical results are limited, and in many cases, one must rely on numerical methods. The Log derivative method is particularly convenient in this context. It was originally introduced for scattering calculations \cite{johnson:1973} and later refined in several works \cite{hamzavi2013approximate,manolopoulos:1986,manolopoulos:1993,hutson:1994}. Its main advantage is that it avoids the direct propagation of the wave function by working instead with its logarithmic derivative, which leads to a more stable numerical procedure \cite{andrade:2026}.

Motivated by recent developments, it is also interesting to consider potentials that arise in other areas of physics. In particular, $\alpha$--attractor models, originally introduced in cosmology in the context of quintessential inflation \cite{salo:2021,dimopoulos2017quintessential}, lead to smooth potential profiles with adjustable parameters and nontrivial asymptotic behavior \cite{zambrano:2025}. From the point of view of quantum scattering, these potentials provide an alternative setting that is worth exploring.

In this work, we study the one--dimensional time--independent Klein--Gordon equation in the presence of an $\alpha$--attractor potential. The reflection coefficient
$\mathcal{R}$ and transmission coefficient $\mathcal{T}$ are computed using the Log derivative method, and special attention is given to the identification of superradiant regimes \cite{brito2015new}. As a first step, the method is tested on the hyperbolic tangent potential, where known results can be used as a reference before applying it to the $\alpha$--attractor case.
\cite{benchiheub2024,ikhdair:2010,johnson:1973,manolopoulos:1986,manolopoulos:1993,hutson:1994,sun:2011}

This work is organized in the following way: 
Sec. \ref{sec_Scattering} is devoted to studying the scattering solutions of the Klein--Gordon equation and its solution through the Log derivative method,  showing the reflection coefficient $\mathcal{R}$ and transmission coefficient $\mathcal{T}$. In Section \ref{sec_Superradiance}, the superradiance phenomenon is discussed. In Section \ref{sec_Potentials}, the potentials are presented, and the application of the Log derivative method for each one is conducted. Section \ref{sec_Results} shows our Results. The plots of the reflection coefficient $\mathcal{R}$ and the transmission coefficient $\mathcal{T}$  for both potentials are presented, where the superradiance phenomenon can be observed. Finally, in section \ref{sec_Conclusions}, we present the conclusions of this work.

\section{Scattering States}
\label{sec_Scattering}

We consider the one dimensional scattering of a relativistic spin--$0$ particle of mass $m$ described by the Klein--Gordon equation in the presence of a $\alpha$--attractor potential  $V(x)$ \cite{salo:2021}. Since the potential does not depend on time, we solve the stationary Klein--Gordon equation  in natural units $\hbar=c=1$ \cite{greiner2000relativistic},

\begin{equation}
\dfrac{\mathrm{d}^2\phi(x)}{\mathrm{d} x^2} + \left\{[E-V(x)]^2 - m^2\right\}\,\phi(x) = 0,
\label{eq:KG}
\end{equation}
where $\phi(x)$ denotes the scalar field, $E$ is the energy eigenvalue of the particle, and $V(x)$ represents the potential under consideration.

For convenience, we define the local relativistic wave number as follows,

\begin{equation}
k^2(x) \equiv \left[E - V(x)\right]^2 - m^2.
\label{eq:kx}
\end{equation}
therefore, Equation \eqref{eq:KG} takes the form,

\begin{equation}
\dfrac{\D^2\phi(x)}{\D x^2} + k^2(x)\,\phi(x) = 0.
\label{eq:KG-sch}
\end{equation}  

The potential $V(x)$ under consideration approaches constant values in the asymptotic regions, $V(x) \to V_L$ as $x \to -\infty$ and $V(x) \to V_R$ as $x \to +\infty$, which allows for scattering states. 

\subsection{The Log Derivative Method}

\bigskip
We obtain the scattering solutions using the Log derivative method \cite{johnson:1973}, which is particularly well suited for numerical scattering problems. Instead of propagating the wave function directly, the method introduces its logarithmic derivative, which, by definition, is, 

\begin{equation}
y(x) = \dfrac{\phi'(x)}{\phi(x)},
\label{eq:log-dev}
\end{equation}  
where prime indicates the derivative with respect to $x$.

Differentiating Eq \eqref{eq:log-dev} with respect to $x$ and using  Eq. \eqref{eq:KG-sch}, we obtain the first order nonlinear Riccati equation \cite{MelendezLugo2025}, 

\begin{equation}
y'(x) = -k^2(x) - y^2(x). 
\label{eq:riccati}
\end{equation}

Since $y(x)$ is invariant under the overall normalization of $\phi(x)$,
the Riccati formulation avoids numerical instabilities associated with
exponentially amplified wave functions~\cite{johnson1977new}. Moreover, in the right asymptotic region $x \to +\infty$, where $V(x) \to V_R$, i.e., far from the step--like potential, the solutions of the Klein--Gordon equation look like plane waves. Thus, we impose the physical scattering condition of a purely outgoing wave, 

\begin{equation}
\phi(x) \sim e^{i k_R x}, \qquad
k_R = \sqrt{(E - V_R)^2 - m^2},
\label{eq:outgoing}
\end{equation}
where the corresponding boundary condition for the Log derivative is, 

\begin{equation}
y(x_R) = i k_R,
\label{eq:bcR}
\end{equation}  
and $x_R$ is chosen sufficiently far into the asymptotic region.  
Then, we use this boundary condition to integrate the Riccati equation given by Eq. \eqref{eq:riccati} backward from $x_R$ toward the left asymptotic region. 

\subsection{Reflection coefficient $\mathcal{R}$ and transmission coefficient $\mathcal{T}$}

\bigskip
We consider the left asymptotic region $(x \to -\infty)$, in which the wave function takes
the standard scattering form,

\begin{equation}
\phi(x) = e^{i k_L x} + R\,e^{-i k_L x},
\qquad
k_L = \sqrt{(E - V_L)^2 - m^2} 
\label{eq:left_asym}
\end{equation} 
where $R$ is the complex reflection amplitude.  
Evaluating the Log derivative of Eq. \eqref{eq:left_asym} at a point $x_L$ in the left asymptotic region and matching it to the numerical value $y(x_L)$, we get, 

\begin{equation}
R = e^{2 i k_L x_L}
\dfrac{i k_L - y(x_L)}{i k_L + y(x_L)}.
\label{eq:Rcoeff}
\end{equation}  

Then, the reflection coefficient $\mathcal{R}$ is given by, 

\begin{equation}
\mathcal{R}(E) = |R|^{2}.
\label{eq:reflection}
\end{equation}

The transmission coefficient $\mathcal{T}$ must be defined in terms of the flux ratio between the transmitted and incident waves. Therefore, it is given by,

\begin{equation}
\mathcal{T}(E) = \dfrac{k_R}{k_L} |T|^2,
\label{eq:transmission}
\end{equation}
where $T$ is the transmission amplitude. In this formulation, the conservation of flux takes the form,

\begin{equation}
\mathcal{R}(E) + \mathcal{T}(E) = 1,
\label{eq:FluxConservation}
\end{equation}
provided that both asymptotic regions support propagating modes.

\subsection{Numerical Implementation}
\label{sec_Numerical}

\medskip
The Riccati equation Eq. \eqref{eq:riccati} is computed using a fourth order Runge--Kutta (RK4) method~\cite{rk4} over a finite spatial interval $[x_{\min}, x_{\max}]$, chosen such that the potential has already reached its asymptotic values at both boundaries. The interval is discretized into $N_{\mathrm{steps}}$ uniform steps. For a fixed energy $E$, the integration starts at the right boundary $x_R$, where the outgoing wave condition is imposed through Eq. \eqref{eq:bcR} with $k_R=\sqrt{(E-V_R)^2-m^2}$. The Riccati equation is then integrated backward from $x_R$ to $x_L$ using a negative step size $h=(x_L-x_R)/N_{\mathrm{steps}}$. Once the numerical value of the logarithmic derivative at the left boundary, $y(x_L)$, is obtained, the reflection amplitude is computed from Eq. \eqref{eq:Rcoeff}, and the reflection coefficient is determined by Eq. \eqref{eq:FluxConservation}.  This procedure is repeated for a set of energies in order to construct the reflection and transmission spectra.


\section{Superradiance}
\label{sec_Superradiance}

\bigskip
For convenience, we denote the asymptotic wave numbers in the left and right regions as,

\begin{equation}
\nu' \equiv k_L, \qquad \mu' \equiv k_R,
\end{equation}
while the dimensionless parameters $\nu = k_L/(2b)$ and $\mu = k_R/(2b)$ are used when comparing with the analytical solution of the hyperbolic tangent potential.

\medskip
The direction of propagation is determined by the group velocity rather than the sign of the wave number. For the Klein--Gordon equation, the group velocity is given by \cite{calogeracos1999history}

\begin{equation}
\dfrac{\mathrm{d}E}{\mathrm{d}k_L} = \dfrac{k_L}{E - V_L},
\label{group_nu}
\end{equation}

\begin{equation}
\dfrac{\mathrm{d}E}{\mathrm{d}k_R} = \dfrac{k_R}{E - V_R}.
\label{group_muu}
\end{equation}

Therefore, the physically allowed solutions are selected by imposing that the group velocity is positive in the incident region, ensuring that the particle propagates from left to right.

For the $\alpha$--attractor potential, we have five different regions, these regions are observed in table \ref{regions}.

\bigskip
\begin{table}[ht]
\begin{center}
\begin{tabular}{c|c|c|c|c}
\hline\hline
$E>V_R+m$       & $\nu'>0$  &  $\nu\in\Re$  & $\mu'>0$  &  $\mu\in\Re$\\
$V_R+m>E>V_R-m$   & $\nu'>0$  &  $\nu\in\Re$  &           &  $\mu\in\Im$\\
$V_R-m>E>V_L+m$  & $\nu'>0$  &  $\nu\in\Re$  & $\mu'<0$  &  $\mu\in\Re$\\
$V_L+m>E>V_L-m$ &           &  $\nu\in\Im$  & $\mu'<0$  &  $\mu\in\Re$\\
$E<V_L-m$      & $\nu'<0$  &  $\nu\in\Re$  & $\mu'<0$  &  $\mu\in\Re$\\
\hline\hline
\end{tabular}
\caption{\label{regions} Regions for $\nu'$ and $\mu'$ defined by the asymptotic limits $V_L$ and $V_R$.}
\end{center}
\end{table}
\medskip

It is important to note that in the regions $V_R+m>E>V_R-m$ and $V_L+m>E>V_L-m$, the dispersion relations $\mu$ and $\nu$ are purely imaginary, and the transmitted wave attenuates, so $\mathcal{R}=1$. In the region $V_R-m>E>V_L+m$, $k_R < 0 \quad \text{and} \quad k_L > 0$ we have that $\mathcal{R}>1$, so superradiance occurs.

\section{Potentials}
\label{sec_Potentials}

\bigskip
In this section, a comparison is made between the analytical solution of the Klein--Gordon equation for the Hyperbolic tangent potential and those obtained using the Log derivative method in order to verify its efficiency. Once this is demonstrated, we proceed to applied the Log derivative method to the  $\alpha$--attractor potential. We also explicitly present  the application of the Log derivative method to each potential.

\subsection{The hyperbolic tangent potential}

\bigskip
The hyperbolic tangent potential   is a step--like potential that has been studied in the context of the Klein--Gordon equation \cite{rojas:2015} and the DKP equation \cite{valladares2023superradiance}. This potential is given by,

\begin{equation}
V(x)=a \tanh\left(b x\right),
\label{Fig_Vtanh}
\end{equation}
where $a$ represents the height of the potential and $b$ gives the smoothness of the curve. The form of the hyperbolic tangent potential is shown in Fig. (\ref{Fig_Vtanh}). 

\begin{figure}[th!]
\begin{center}
\includegraphics[scale=0.4]{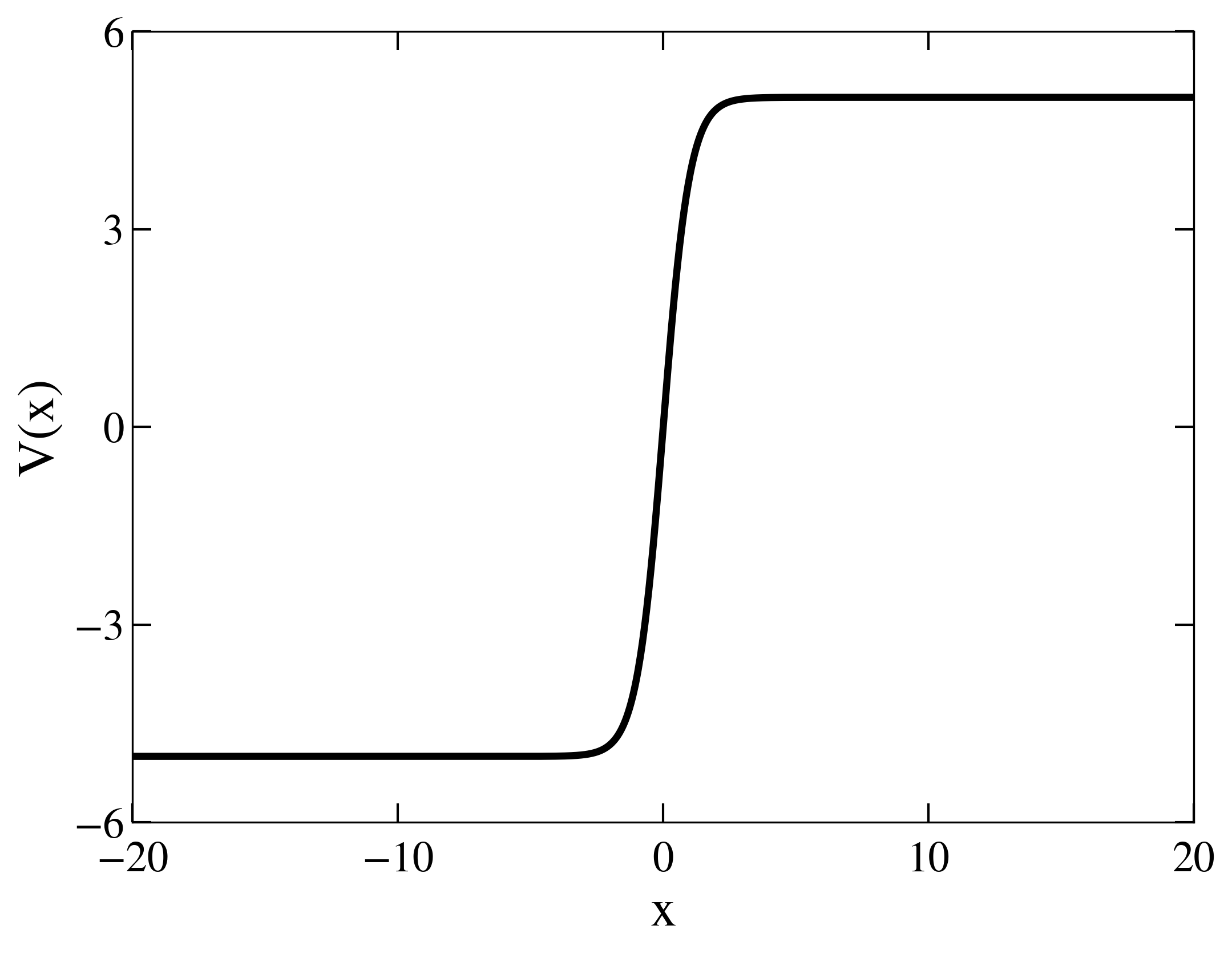}
\caption{The hyperbolic tangent potential  for $a=5$, and $b=1$.}
\label{fig_Vtanh}
\end{center}
\end{figure}

From Fig. \ref{fig_Vtanh}, it is observed that the hyperbolic tangent potential approaches a step--like profile in the limit $b \rightarrow \infty$. Using the asymptotic behavior of the hyperbolic tangent function, we obtain

\begin{equation}
V_L = \lim_{x \to -\infty} V(x) = -a, \qquad
V_R = \lim_{x \to +\infty} V(x) = a.
\end{equation}

Therefore, the relativistic energy thresholds are given by $E = V_L \pm m = -a \pm m$ and $E = V_R \pm m = a \pm m$. These values determine the different scattering regimes and are used to label the energy axis in the corresponding figures.

The analytical solution for the Hyperbolic tangent potential was presented by Rojas \cite{rojas:2015}, where, after solving the Klein–Gordon equation in terms of the hypergeometric functions ${}_2F_1(a,b;c;z)$ \cite{abramowitz:1965}, the reflection and transmission coefficients are obtained  using the definition of the electrical current density for the one--dimensional time--independent Klein--Gordon equation,

\begin{eqnarray}
\label{R}
\mathcal{R}&=&\dfrac{|B|^2}{|A|^2},\\
\mathcal{T}&=&\dfrac{\mu}{\nu}\frac{1}{|A|^2},
\end{eqnarray}
with,

\begin{eqnarray}
\label{A}
A&=&\frac{\Gamma(1-2i\mu)\Gamma(-2i\nu)}{\Gamma(-i\nu+\lambda-i\mu)\Gamma(1-i\nu-\lambda-i\mu)},\\
\label{B}
B&=&\frac{\Gamma(1-2i\mu)\Gamma(2i\nu)}{\Gamma(i\nu-\lambda-i\mu)\Gamma(1+i\nu-\lambda-i\mu)},
\end{eqnarray}
where

\begin{eqnarray}
\label{alpha}
\nu&=&\dfrac{\sqrt{(E+a)^2-m^2}}{2b},\\
\label{beta}
\lambda&=&\dfrac{b+\sqrt{b^2-4a^2}}{2b},\\
\label{gamma}
\mu&=&\dfrac{\sqrt{(E-a)^2-m^2}}{2b}.
\end{eqnarray}

\bigskip
The reflection coefficient $\mathcal{R}$ and the transmission coefficient $\mathcal{T}$ are expressed in terms of the coefficients $A$ and $B$, therefore, they are expressed in terms of the Gamma function and are determined using the software Wolfram Mathematica 14.3 \cite{Mathematica}. This potential presents superradiance, the regions are defined in the Figures \ref{fig_Rtanh} and \ref{fig_Ttanh}.

\subsection{The $\alpha$--attractor potential}

\begin{figure}[th!]
\begin{center}
\includegraphics[scale=0.4]{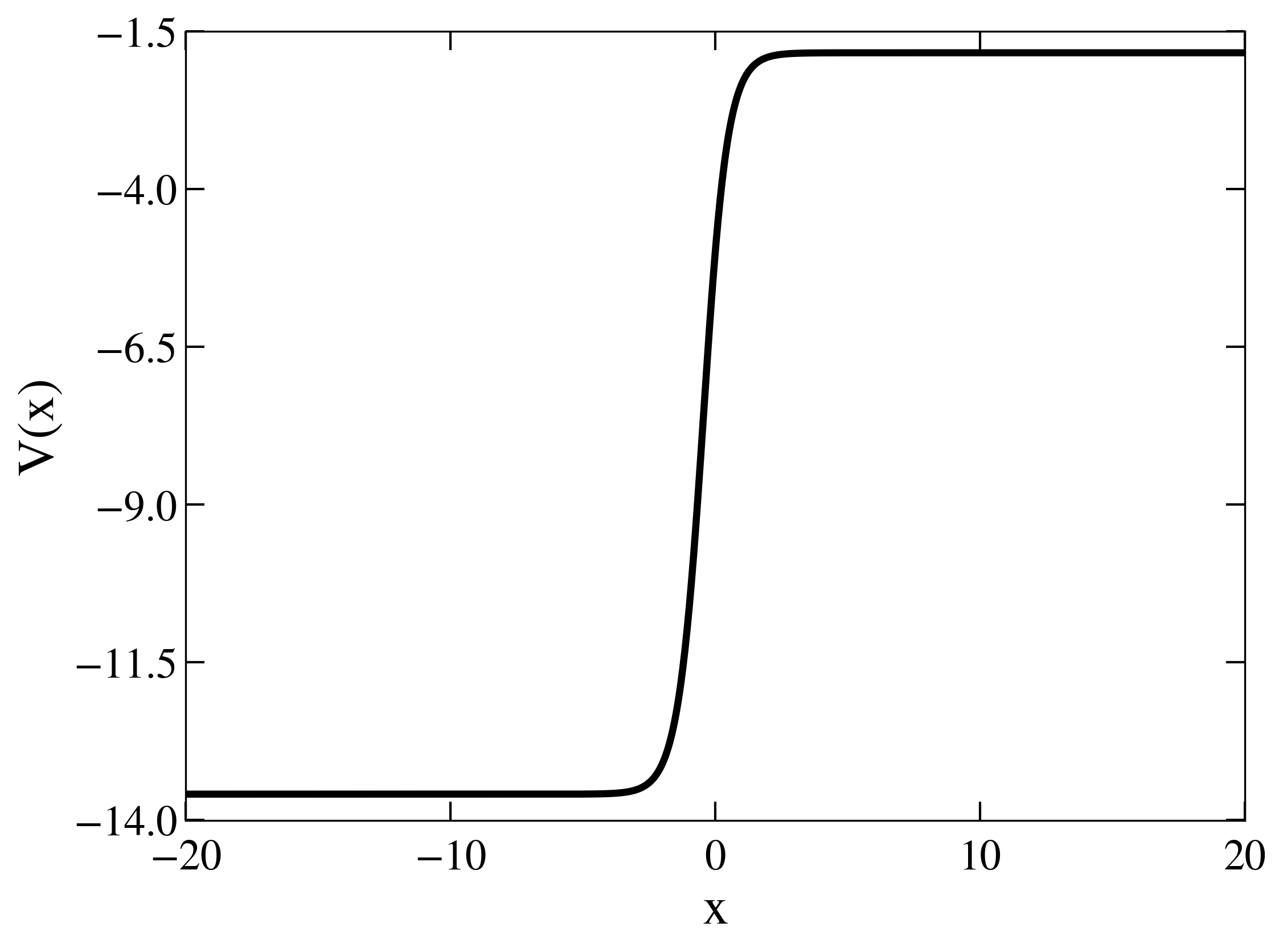}
\caption{The $\alpha$--attractor potential  for $a=-5$, $b=1$, and $c=1$.}
\label{Fig_Valpha}
\end{center}
\end{figure}

\bigskip
The $\alpha$--attractor potential   is an asymmetric step--like potential, and it is  defined by Sal\'o \cite{salo:2021} in the context of quintessential inflation as,

\begin{equation}
V(x)=a \, e^{-b\tanh\left( c x\right)},
\label{Valpha}
\end{equation}
where $a$ represents the height of the potential and $c$ represents the smoothness of the potential. The parameter $b$ changes the amplitude of the potential. This potential is represented in Fig. \eqref{Fig_Valpha}. Potential $\alpha$--attractor  can be compared with the hyperbolic tangent potential  by a specific relationship among its parameters. Additionally, note that when $c \rightarrow \infty$ the $\alpha$--attractor potential   reduces to a step potential.

To better understand the scattering properties of the $\alpha$--attractor potential, it is useful to analyze its asymptotic behavior. From Eq. \eqref{Valpha}, and using the limits of the hyperbolic tangent function,

\begin{equation}
\lim_{x \to -\infty} \tanh(cx) = -1, \quad \lim_{x \to +\infty} \tanh(cx) = +1,
\end{equation}
the potential approaches constant values at both spatial infinities,

\begin{equation}
V_L = \lim_{x \to -\infty} V(x) = a e^{b}, \quad
V_R = \lim_{x \to +\infty} V(x) = a e^{-b},
\end{equation}
it shows that the $\alpha$--attractor potential behaves as an asymmetric step--like potential connecting two different constant regions.

This structure is directly analogous to the relativistic step potential associated with the Klein paradox. The asymptotic regions define the effective relativistic energy thresholds $E = V_L \pm m$ and $E = V_R \pm m$, which separate different scattering regimes. These thresholds determine whether the wave numbers $k_L$ and $k_R$ are real or imaginary, and therefore whether propagation or evanescent behavior occurs.

Therefore, the asymptotic structure of the $\alpha$--attractor potential provides a clear physical interpretation. By substituting the limits $V_L = a e^b$ and $V_R = a e^{-b}$ into the general regimes defined in Table \ref{regions}, the specific energy bands for total reflection and superradiance in this asymmetric potential are fully determined.

\subsection{Application of the Log Derivative Method}

\bigskip
In this section, the numerical procedure described in Section~\ref{sec_Numerical} is applied to both the hyperbolic tangent potential and the $\alpha$--attractor potential by specifying the corresponding functional form of $V(x)$ and its asymptotic limits.

\medskip
For the hyperbolic tangent potential defined in Eq.~\eqref{Fig_Vtanh}, the asymptotic values are $V_L = -a$ and $V_R = a$. Therefore, the local wave number entering the Riccati equation becomes,

\begin{equation}
k^2(x) = [E - a \tanh(bx)]^2 - m^2.
\end{equation}

The boundary condition at $x_R$ is imposed using,

\begin{equation}
k_R = \sqrt{(E - V_R)^2 - m^2} 
= \sqrt{(E - a)^2 - m^2}.
\end{equation}

\medskip
For the $\alpha$--attractor potential defined in Eq.~\eqref{Valpha}, the asymptotic limits are $V_L = a e^{b}$ and $V_R = a e^{-b}$. The corresponding local wave number is given by,

\begin{equation}
k^2(x) = [E - a e^{-b\tanh(cx)}]^2 - m^2.
\end{equation}

The boundary condition is imposed using,

\begin{equation}
k_R = \sqrt{(E - V_R)^2 - m^2} = \sqrt{(E - ae^{-b})^{2} - m^{2}}.
\end{equation}

Once the potential and its asymptotic limits are fully specified, the Riccati equation is integrated over the chosen spatial domain, allowing the reflection coefficient $\mathcal{R}$ and transmission  coefficient $\mathcal{T}$ to be computed. 

\section{Results}
\label{sec_Results}

\bigskip
\subsection{The hyperbolic tangent potential}

\begin{figure}[htbp]
\begin{center}
\includegraphics[scale=0.40]{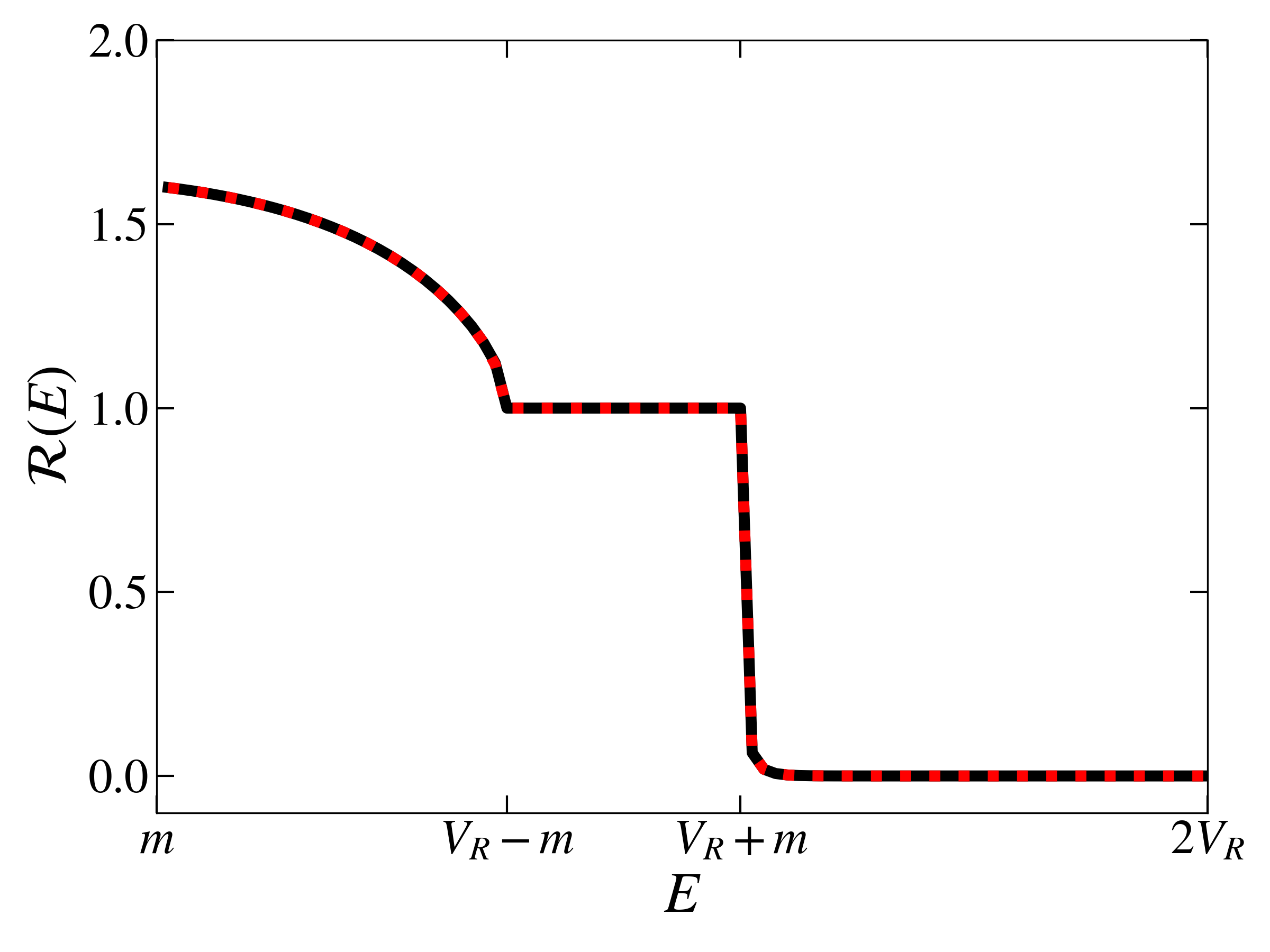}
\caption{The reflection coefficient $\mathcal{R}$ for the hyperbolic tangent potential with $a=5$ and $b=1$. The asymptotic limits are $V_L = -a$ and $V_R = a$, so the energy thresholds correspond to $E = V_R \pm m$. Solid black line: analytical solution, dashed red line: Log derivative method.}
\label{fig_Rtanh}
\end{center}
\end{figure}

\begin{figure}[htbp]
\begin{center}
\includegraphics[scale=0.40]{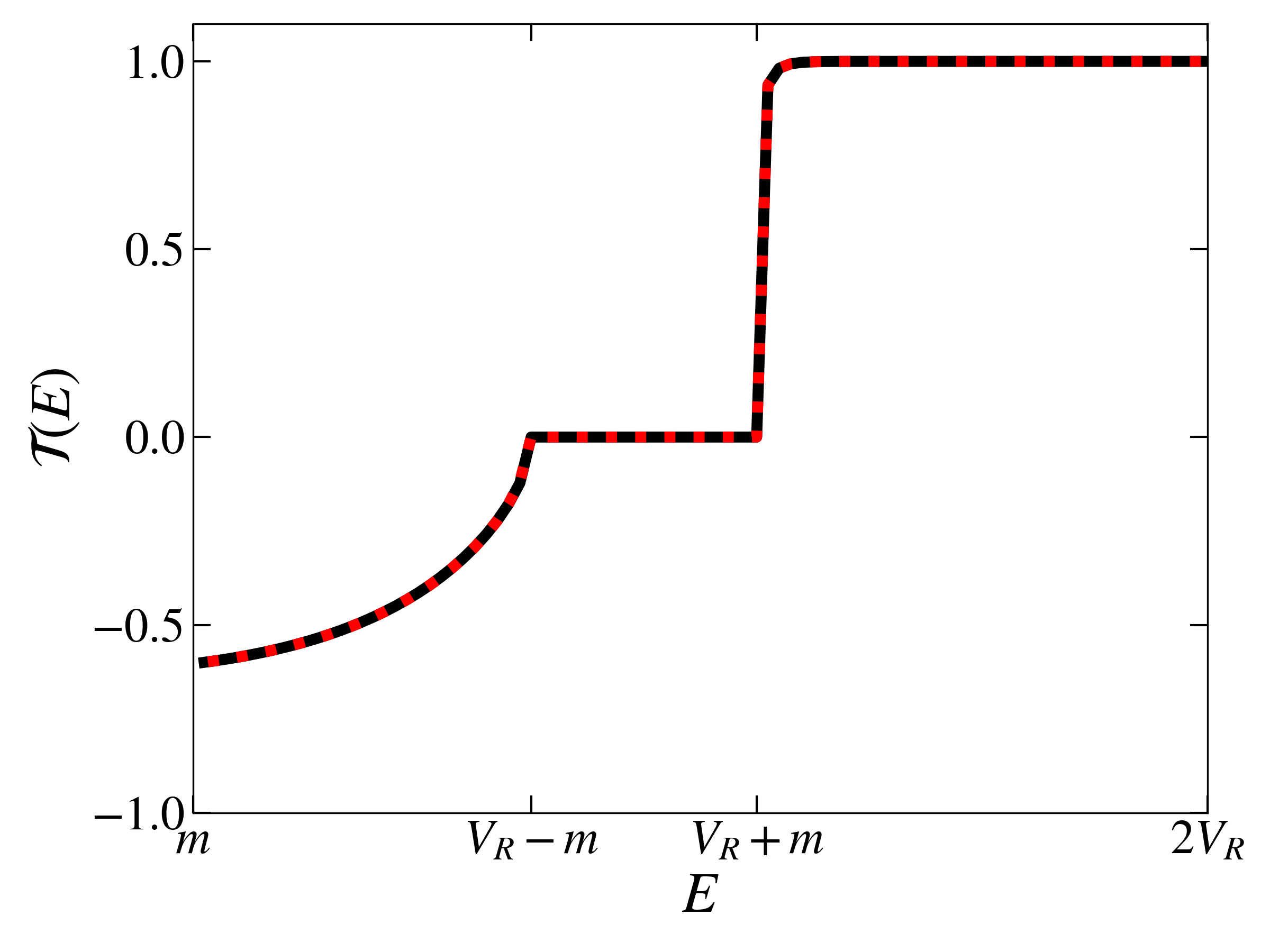}
\caption{The transmission coefficient $\mathcal{T}$ for the hyperbolic tangent potential with $a=5$ and $b=1$. The asymptotic limits are $V_L = -a$ and $V_R = a$, so the energy thresholds correspond to $E = V_R \pm m$. Solid black line: analytical solution, dashed red line: Log derivative method.}
\label{fig_Ttanh}
\end{center}
\end{figure}

\bigskip
In Figs. \ref{fig_Rtanh} and \ref{fig_Ttanh}, the plots for the reflection  coefficient  $\mathcal{R}$ and the transmission coefficient $\mathcal{T}$  versus the energy $E$ are shown, calculated analytically and using the Log derivative method for the parameters $a=5$ and $b=1$. For all numerical computations, we set $m = 1$. We can conclude that the Log derivative method is a useful method to calculate scattering states and study the superradiance phenomenon.

\subsection{The $\alpha$--attractor potential}

\begin{figure}[th!]
\begin{center}
\includegraphics[scale=0.40]{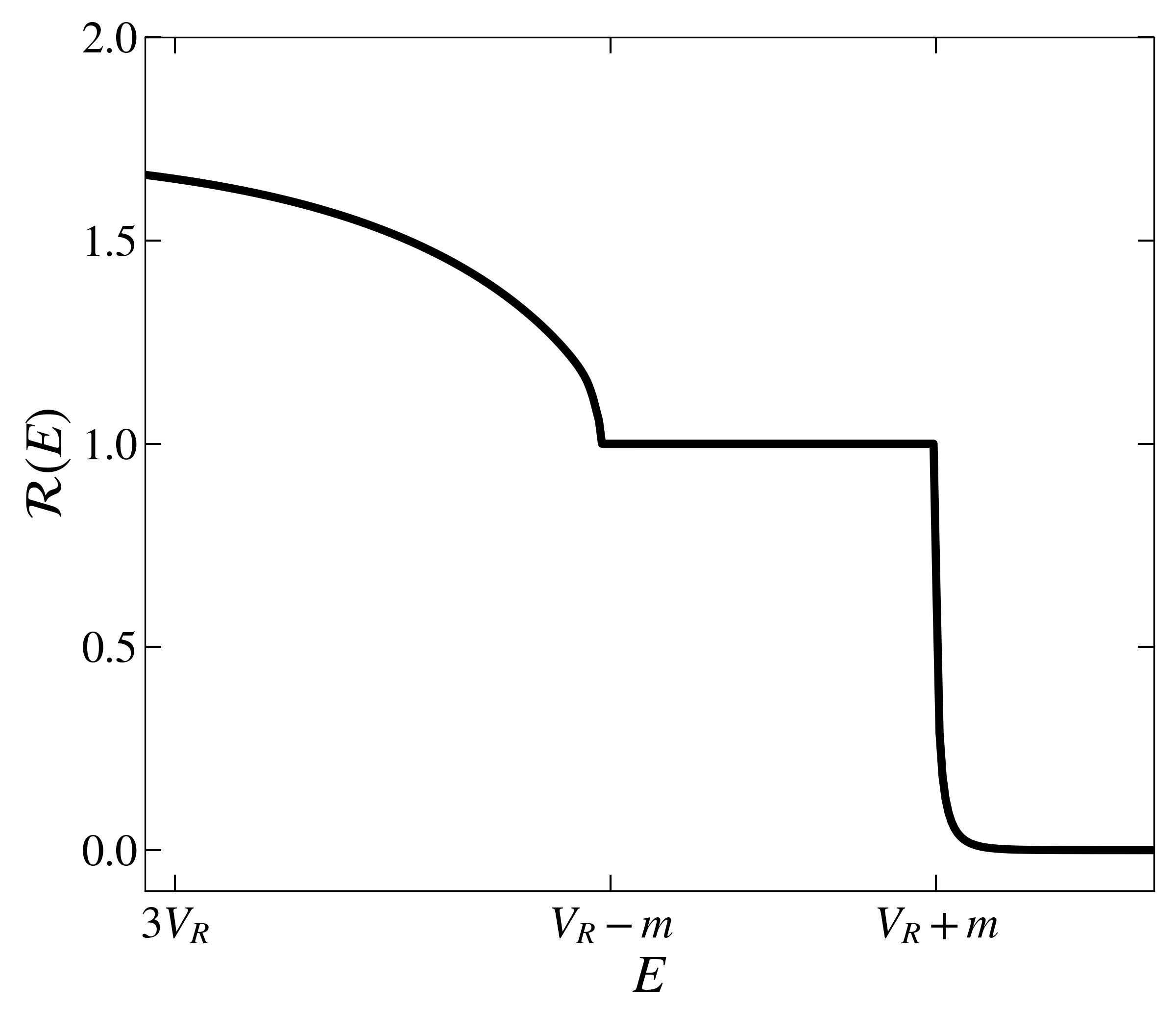}
\caption{The reflection coefficient $\mathcal{R}$ for the $\alpha$--attractor potential with $a=-5$, $b=1$, and $c=1$, computed using the Log derivative method. The asymptotic limits are $V_L = a e^{b}$ and $V_R = a e^{-b}$, which determine the corresponding energy thresholds.}
\label{fig_Ralpha}
\end{center}
\end{figure}

\begin{figure}[th!]
\begin{center}
\includegraphics[scale=0.40]{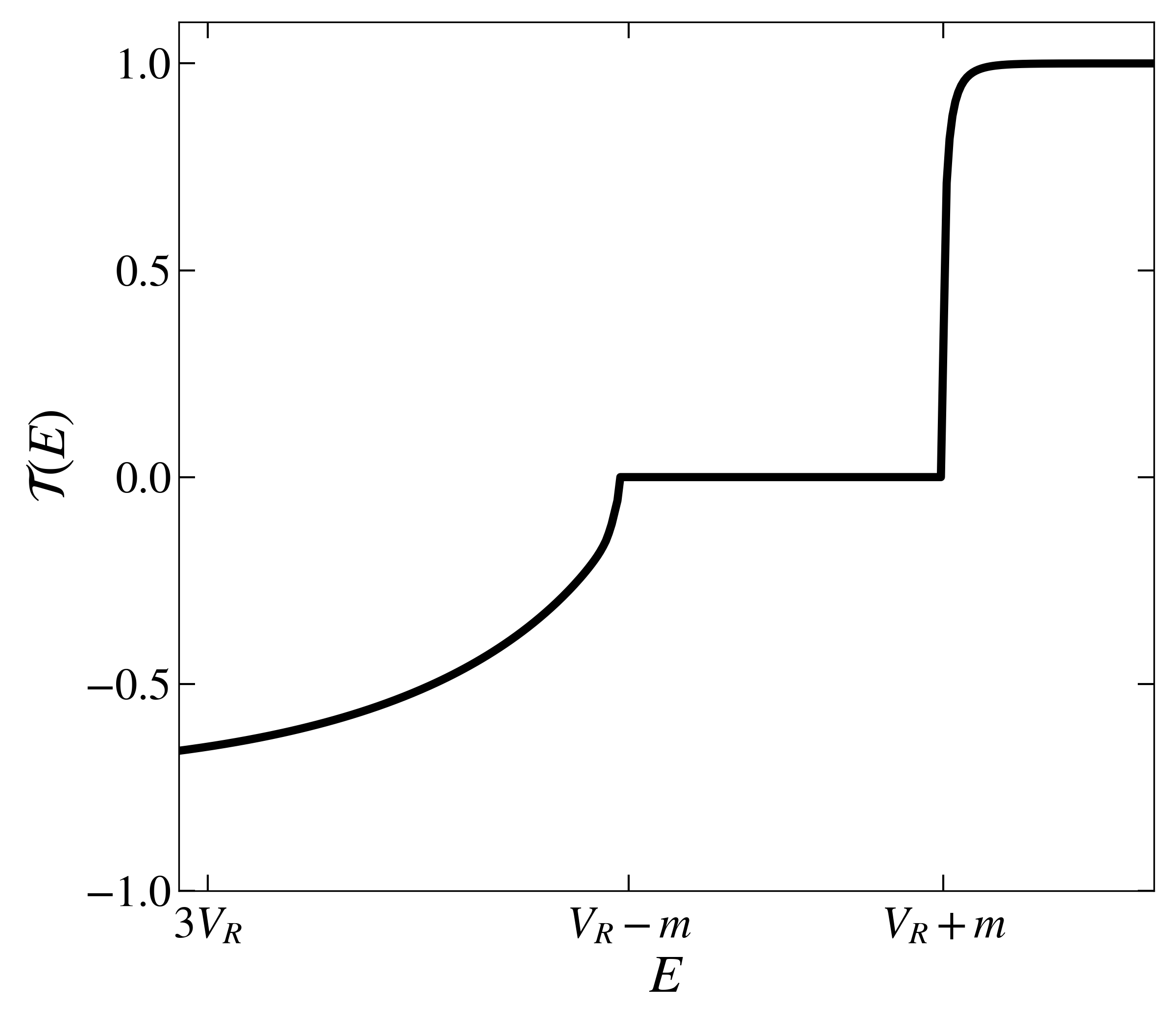}
\caption{The transmission coefficient $\mathcal{T}$ for the $\alpha$--attractor potential with $a=-5$, $b=1$, and $c=1$, computed using the Log derivative method. The asymptotic limits are $V_L = a e^{b}$ and $V_R = a e^{-b}$, which determine the corresponding energy thresholds.}
\label{fig_Talpha}
\end{center}
\end{figure}

\bigskip
In Figs. \ref{fig_Ralpha} and \ref{fig_Talpha}, the reflection coefficient $\mathcal{R}$  and the transmission coefficient  $\mathcal{T}$ are presented for the $\alpha$--attractor potential as functions of the energy $E$ and computed using the Log derivative method with the parameters $a=-5$, $b=1$, and $c=1$. In all cases, the mass is fixed at $m=1$.

As shown in Fig. \ref{fig_Ralpha}, the numerical results are fully consistent with the theoretical predictions derived from the asymptotic analysis of the potential. The total reflection region ($\mathcal{R}=1$) appears in the energy interval where the transmitted wave becomes evanescent, namely $V_R + m > E > V_R - m$. 

More importantly, the superradiance regime is clearly observed in the interval $V_R - m > E > V_L + m$, where the reflection coefficient exceeds unity ($\mathcal{R} > 1$). In this region, the transmitted wave corresponds to states with negative effective energy, which leads to an amplification of the reflected flux. 

These results demonstrate that the Log derivative method accurately captures both the transition between propagating and evanescent regimes and the onset of superradiance, even in the presence of a strongly asymmetric potential.

\section{Conclusions}
\label{sec_Conclusions}

\bigskip
In this work, the time--independent one--dimensional Klein--Gordon equation is solved in the presence of the $\alpha$--attractor potential through the Log derivative method. The reflection coefficient $\mathcal{R}$ and  the transmission  coefficient $\mathcal{T}$ are calculated, and the superradiance phenomenon is observed.  By comparing these coefficients with the corresponding analytical results for the hyperbolic tangent potential, we have shown that the logarithmic derivative method yields exact solutions, thus confirming the reliability of its predictions for the $\alpha$--attractor potential.

\section{Acknowledments}

\bigskip
\noindent
L. M. P. acknowledges financial support from ANID through Convocatoria Na-
cional Subvenci\'on a Instalaci\'on en la Academia Convocatoria A\~no 2021, Grant No.
SA77210040.

\bigskip
\noindent
D. L. acknowledges partial financial support from Centers of Excellence with
BASAL/ANID financing, Grant No. AFB220001, CEDENNA


\end{document}